\documentclass[aps,showpacs,floatfix,twocolumn,amsmath,amssymb,superscriptaddress,prd]{revtex4-1}
\usepackage{tikz,graphicx,microtype}
\usepackage{booktabs}
\usepackage{multirow}

\newcommand{\Ef}{\mathcal{E}}

\newcommand{\EA}{\text{EN1}}
\newcommand{\EB}{\text{EN2}}
\newcommand{\EC}{\text{EN3}}
\newcommand{\ED}{\text{EN4}}

\newcommand{\EEA}{\text{EN5}}
\newcommand{\EEB}{\text{EN6}}
\newcommand{\EEC}{\text{EN7}}
\newcommand{\EED}{\text{EN8}}

\usepackage[colorlinks=true,backref=false, linktocpage=true,
citecolor=blue,urlcolor=blue,linkcolor=blue,pdfpagemode=UseOutlines]{hyperref}

\hypersetup{%
  bookmarksnumbered=true,
  pdftitle = {},
  pdfsubject = {},
  pdfauthor = {A.~Alexandru},
  pdfkeywords = {}
}

\def\fm {\,{\rm fm}}
\def\MeV {\,{\rm MeV}}

\begin{document}

\title{Finite volume effects on the electric polarizability of neutral hadrons in lattice QCD}
\author{M. Lujan}
\affiliation{Math Modeling Group, LM$\text{$\tilde{I}$}$, Tysons Corner, VA 22102, USA}
\author{A. Alexandru}
\affiliation{Department of Physics, The George Washington University, Washington, DC 20052, USA}
\affiliation{Department of Physics, University of Maryland, College Park, MD 20742}
\author{W. Freeman}
\affiliation{Department of Physics, Syracuse University,  Syracuse, NY 13244, USA}
\author{F.X. Lee}
\affiliation{Department of Physics, The George Washington University, Washington, DC 20052, USA}

\begin{abstract}
We study the finite volume effects on the electric polarizability for the neutron, neutral pion, and neutral kaon
using eight dynamically generated two-flavor nHYP-clover ensembles at two different pion masses: 306(1) and 227(2) MeV. 
An infinite volume extrapolation is performed for each hadron at both pion masses.
For the neutral kaon, finite volume effects are relatively mild. The dependence on the quark mass is also mild and 
a reliable chiral extrapolation can be performed along with the infinite volume extrapolation. 
Our result is $\alpha_{K^0}^\text{phys}=0.356(74)(46) \times 10^{-4} \fm^3$. 
In contrast, for neutron the electric polarizability depends strongly on the volume. After removing the finite volume corrections, 
our neutron polarizability results are in good agreement with $\chi$PT.
For the connected part of the neutral pion polarizability, the negative trend persists, and it is not due to finite volume effects, 
but likely sea quark charging effects.

\end{abstract} 
\pacs{12.38.Gc}

\maketitle

\section{Introduction} \label{secintroduction}

Hadron electromagnetic polarizabilities encode important information about the distribution of charge and current densities
inside the hadrons. Experimentally these parameters are extracted using cross-sections measured in Compton scattering reactions with theoretical input from effective models and dispersion relations. Lattice QCD can provide 
first-principles-based results for static polarizabilities directly as predicted by quark-gluon dynamics. 
This input is particularly important for unstable
hadrons, where experimental and theoretical uncertainties in the effective models are large. 

At the lowest order the effects of an electromagnetic field on hadrons can be parameterized by the effective Hamiltonian:
\begin{equation}
\mathcal{H}_{em} = -\vec{p}\cdot\vec{\Ef} -\vec{\mu}\cdot\vec{B} -\frac{1}{2}\left(\alpha \Ef^2 + \beta B^2\right)+\cdots,
\label{eq:1}
\end{equation}
where $p$ and $\mu$ are the static electric and magnetic dipole moments, respectively, 
and $\alpha$ and $\beta$ are the static electric and magnetic polarizabilities. 
Due to time reversal symmetry of the strong interaction, the static dipole moment, 
$\vec{p}$, vanishes.  In the presence of a constant electric field only, 
the leading contribution to the electromagnetic interaction comes from the electric 
polarizability term at $\mathcal{O}(\Ef^2)$.

Lattice QCD calculations of electromagnetic polarizabilities are challenging since the electromagnetic effects are small 
compared to the natural hadronic scale. A good understanding of all systematic effects is required to ensure that the 
parameters extracted from these calculations are reliable. To that end, our first goal was to validate our method by 
focusing on the neutron electric polarizability. For neutral hadrons lattice QCD calculations are more reliable than for charged hadrons since 
neutral particles are not accelerated by the external field. On the experimental side results for the neutron are reasonably 
precise and effective model predictions are in good agreement with the experimental data. This makes 
the lattice QCD extraction of the electric polarizability of the neutron a good benchmark study.  

In a previous study we computed the electric polarizability of the neutron, neutral pion, and neutral kaon for two 
different pion masses (306 and 227~MeV) with a fixed box size of $L \simeq 3 \fm$~\cite{Lujan:2014kia}. The results 
we found were a bit puzzling: the pion polarizability exhibited the same negative trend observed in other studies 
both with dynamical~\cite{Detmold:2009dx} and quenched ensembles~\cite{Alexandru:2010dx} and the neutron polarizability 
was in disagreement with predictions from chiral perturbation 
theory~\cite{Lensky:2009uv,McGovern:2012ew,Griesshammer:2012we,Griesshammer:2015ahu}. 
We speculated that corrections due to electrically neutral sea quarks or finite-volume effects
could explain these discrepancies. A calculation of the polarizability, with the inclusion of the charged sea quarks, 
was done on the 306~MeV ensemble~\cite{Freeman:2014kka,Freeman:2014gsa}. It was found that charging the sea 
quarks does not change the polarizability significantly, which is aligned with expectations from chiral perturbation theory.
Thus, the discrepancy between our lattice calculation of the neutron polarizability and the calculation from 
$\chi$PT remained. In this paper we study the finite-volume corrections for this quantity.

The paper is organized as follows: In Section~\ref{secmethodology} we present the method used to extract 
the polarizability from  the lattice for mesons and baryons. This includes a discussion of our fitting procedure. 
In Sections~\ref{sec:voldependence} we present our results of the polarizability for the neutron, pion, and kaon 
and discuss the finite-volume corrections. In Section~\ref{sec:discussion} we discuss the quark mass dependence for
the infinite volume extrapolated polarizability and compare our results with predictions from $\chi$PT.
Lastly, in Section~\ref{sec:conclusion} we summarize our results and outline 
our plans for future investigations. 

\bigskip

\section{ Methodology} \label{secmethodology}


\subsection{Background field method} \label{seceshift}

In lattice QCD polarizabilities can be computed using the background field method~\cite{Martinelli:1982cb}: 
the energy shift induced by a constant electric field is directly related to the static electric polarizability.
A static electromagnetic field can be introduced by coupling the vector potential ($A_{\mu}$) to 
the covariant derivative of the Euclidean QCD Lagrangian,
\begin{equation}
D_{\mu} = \partial_{\mu} -igG_{\mu} -iqA_{\mu},
\end{equation}
where $G_{\mu}$ is the gluon field. On the lattice this is implemented by a multiplicative 
U(1) phase factor to the gauge links {\it{i.e.}},
\begin{equation}
U_{\mu} \rightarrow e^{-iqaA_{\mu}} U_{\mu}.
\end{equation}

For a constant electric field, one choice for the vector potential is $A_x = \Ef t$, where we have used an 
imaginary value for the electric field leading to a U(1) multiplicative factor that keeps the links unitary. 
When using an imaginary value of the field, the energy shift due to the polarizability acquires an additional 
negative sign so that a positive energy shift corresponds to a positive value of the polarizability~\cite{Alexandru:2008sj}. 

In this study we use very weak electric fields to extract the polarizability, 
so that the energy shift is due to polarizabilities rather than higher order terms in the effective 
Hamiltonian in Eq.~\ref{eq:1}.
It is possible of course to extract the polarizability using stronger fields, but this
would require the evaluation of the energy shifts for different electric field strengths to reliably separate
the higher order corrections. 
We use Dirichlet boundary conditions (DBC) for the valence quarks in both the time direction and the direction of 
the electric field. This choice of boundary conditions allows us to chose an arbitrarily small value of the electric field.
In our analysis we use a value of
\begin{equation}
\eta \equiv a^2 q_d\Ef = 10^{-4},
\end{equation} 
where $a$ is the lattice spacing and $q_d$ is the magnitude of the electric charge for the down quark. 
One bound on $\eta$ is determined by looking at a $\pm{\cal E}$-symmetrized hadron correlator (see below) at various time slices
and determining the range of $\eta$ values which exhibit quadratic scaling~\cite{Lujan:2014kia}. 
A more stringent constraint on $\eta$ appears when we take into account the
effect of the sea-quark charge via perturbative reweighting~\cite{Freeman:2014kka}. The latter constraint forces us
to use this low $\eta$ value.
In physical terms, this value of $\eta$ corresponds to an electric field that an electron would generate at a 
distance of $0.5\fm$. The value is well within the quadratic scaling region. Note that the value is about $50$
times lower than the lowest quantized value $2\pi/(N_x \times N_t)$ corresponding to one unit of electric flux;
thus the induced energy shift is thousands of times smaller. 
In our study, the energy shift for the neutron is on the order of keV out of 938 MeV.

Due to the boundary conditions, the quark and hadrons' correlators
close to the boundaries will behave differently than in the bulk.
These effect are enhanced when the source is placed close to the
walls (see for example the discussion about correlators with    
sources close to discontinuities in the non-quantized      
background fields~\cite{Detmold:2009dx,Davoudi:2015cba}.) To minimize these effects, we placed
the source for our quark correlators at maximal distance from the
spatial walls and six lattice units from the temporal wall. In any case, the hadron propagator
will be affected by the walls since the particle in the lowest
momentum state will have a non-zero probability to be within
the distortion region due to the hard walls. Since this        
region is expected to have finite range, the corrections will
be proportional with the probability to be in this region, which
is will vanish as we increase the distance $L$ between the walls
as $1/L$ (recall that we only have hard walls in one spatial direction.)
These corrections will appear as a finite volume correction.

To determine the energy shift $\delta E$ on the lattice we calculate the zero-field ($G_0$), plus-field ($G_{+\Ef}$),  and minus-field ($G_{-\Ef}$) two-point correlation functions for the interpolating operators of interest. The combination of the plus and minus field correlators allows us to remove any $\mathcal{O}(\Ef)$ effects, which are statistical artifacts, when the sea quarks are not charged. For neutral particles in a constant electric field the correlation functions still retain their single exponential decay in the limit $t \rightarrow \infty$,
\begin{equation}
\langle G_{\Ef}(t) \rangle \underset{t\to\infty}{\approx} A(\Ef)e^{-E(\Ef)t}, 
\label{eqn::corr}
\end{equation}
where $E(\Ef)$ has the perturbative expansion in the electric field given by
\begin{equation}\label{eqn.energy}
E(\Ef) = m + \frac{1}{2}\alpha \Ef^2 + ... ~.
\end{equation}
By studying the variations of the correlation functions with and without an electric field one can isolate the energy shift to obtain $\alpha$.

For spin-1/2 hadrons, the energy shift in a constant electric field receives a contribution
due to the magnetic moment of the hadron at order ${\cal O}({\cal E}^2)$. Thus the static polarizability $\alpha$ defined by Eq.~\ref{eqn.energy},
is not identical to the Compton polarizability $\bar{\alpha}$ that enters the effective Lagrangian for spin-1/2 
systems~\cite{Lvov:1993fp}. The relation between these polarizabilities can be computed~\cite{Detmold:2009dx,Lujan:2014kia}.
For these systems the energy expansion reads,
\begin{equation}
E(\Ef) = m + \frac{1}{2}\Ef^2\left(\bar{\alpha}- \frac{\mu^2}{m}\right) + ...~,
\label{eqn::Eneutron}
\end{equation}
where $\bar{\alpha}$ is the Compton polarizability that we wish to compute. To account for the magnetic moment we use the same procedure as we did in a previous study~\cite{Lujan:2014kia}. 

Since we use Dirichlet boundary conditions, the lowest energy state corresponds to a hadron moving with a momentum roughly
equal to $\pi/L$, which vanishes in the limit $L\rightarrow \infty$. When we extract the energy shift from the
hadron we need to account for the induced momentum because the energy shift ($\delta E$) is not equal to the mass 
shift ($\delta m$). The two are related via  the dispersion relation $E = \sqrt{m^2 + p^2}$ by 
\begin{equation}\label{eqn:dEtodm}
\delta m = \delta E \frac{E}{m},
\end{equation}
where $m$ is the zero-momentum mass of the particle which we calculate using periodic boundary conditions (PBC). The mass shift 
$\delta m$ is then used in
Eq.~\ref{eqn.energy} or Eq.~\ref{eqn::Eneutron}, to extract the polarizability.


\subsection{Fitting Method}
\label{fitmethod}

Since the correlation functions $G_0, G_{+{\cal E}}$, and $G_{-{\cal E}}$ are dominated by a single exponential at large times,  
we can use standard spectroscopy techniques to measure the shift in a hadron's energy. 
The only caveat is that the shift is very small at the field strength used in this study, smaller than the statistical 
errors if they were fitted separately.
To overcome the difficulty,  we take advantage of the fact that the three correlators are highly correlated since they are computed on the same set of gauge configurations.  
To do this we construct the combined residue vector from the individual residue vectors in each sector,
\begin{eqnarray}
\mathbf{v}_{i} &\equiv& f(t_i) - \langle G_{0}(t_i)\rangle, \nonumber\\
\mathbf{v}_{N+i} &\equiv&  \bar{f}(t_{i}) - \langle G_{+\Ef}(t_{i})\rangle, \\
\mathbf{v}_{2N+i} &\equiv&  \bar{f}(t_{i}) - \langle G_{-\Ef}(t_{i})\rangle, \nonumber
\end{eqnarray}
where  $i=1,\cdots,N$ labels the time slices in the fit window, $f(t) = A~e^{-E t}$ the fitting function in the 
absence of the field,
 and $\bar{f}(t) =(A+\delta A)~e^{-(E+\delta E)t}$ the fitting function in the 
presence of the field.
We minimize the $\chi^2$ function, 
\begin{equation}
\chi^2= \mathbf{v}^T \mathbf{C}^{-1} \mathbf{v}, 
\end{equation}
for four parameters ($A$, $E$, $\delta A$, $\delta E$) in the usual fashion, where $\mathbf{C}$ is the $3N\times 3N$ jackknifed covariance matrix which takes into account the correlations both in time and in the electric field.
Specifically, the matrix has a $3\times 3$ block structure
\[ \mathbf{C} = \left( \begin{array}{ccc}
C_{0 0} ~& C_{0 +}~ & C_{0 -} \\
C_{+ 0} ~& C_{+ +}~ & C_{+ -} \\
C_{- 0} ~& C_{- +} ~& C_{- -} \end{array} \right),\]
where $0,+,-$ represent  $G_0, G_{+\Ef}$, and $G_{-\Ef}$ respectively.  Each block is a $N\times N$ matrix.
The correlations are encoded in the off-diagonal blocks. 
Note that the symmetrization in the electric field is done implicitly in this procedure, 
since $\bar{f}$ is the same for  $G_{+\Ef}$, and $G_{-\Ef}$. 
 The statistical errors on the parameters are derived from the Hessian of the $\chi^2$.
 This method is used to extract all parameters presented in this work.

To illustrate the importance of accounting for these correlations, we consider the energy shift $\delta E$ for the neutron 
for one of the ensembles used in this work. Using the full covariance matrix we find 
$a\,\delta E = (4.3\pm1.2) \times 10^{-7}$. If we neglect the correlations, which is equivalent to using only 
the diagonal blocks of the covariance matrix, we find $a\, \delta E = (8.15 \pm 150000) \times 10^{-7}$, 
which has huge errors.


\subsection{Calculation details}
\label{sec:ensemble}

We calculate the electric polarizability for the neutron, neutral pion, and neutral kaon on eight dynamically 
generated ensembles using 2-flavor nHYP-clover fermions~\cite{Hasenfratz:2007rf}. For the neutral pion polarizability we are computing 
only the connected contribution to the pion correlation function, as we had 
also done in~\cite{Lujan:2014kia}. We used two quark masses, 
corresponding to pion masses of $227(2)\MeV$  and $306(1)\MeV$. 
For each mass we performed simulations on four different volumes, to study finite volume effects.
 
To save time, we varied the dimension of the lattice only along the electric field ($x$-direction). 
We expect that the finite-volume corrections vanish exponentially in the transverse directions, and that our
lattice is large enough for these corrections to be negligible at the current precision level. On the other
hand, the corrections associated with the direction parallel with the electric field are expected to vanish
only as a power law in $1/L$. We will show that our results agree with these expectations. 

Details of the ensembles are given in Table~\ref{tab:ensembles}. The determination of both the lattice spacing 
and $\kappa_s$, the hopping parameter for the strange quark that is required to compute the kaon polarizability, 
is discussed in detail in our previous study~\cite{Lujan:2014kia}. We use the same values here: 
$\kappa_s=0.1266$ for ensembles EN1 to EN4; $\kappa_s=0.1255$ for ensembles EN5 to EN8.

%
\begin{table}[t]
\begin{tabular*}{0.9\columnwidth}{@{\extracolsep{\stretch{1}}}*{6}c@{}}
\toprule
Label &Lattice&  $a$ (fm) &  $\kappa$& $N_{\text{c}}$& $N_{s}$\\ 
\midrule
$\EA$&$16\times16^2 \times 32$  & 0.1245& 0.12820  &  230&11\\
$\EB$&$24\times 24^2 \times 48$ & 0.1245& 0.12820  &  300& 25\\
$\EC$&$30 \times 24^2 \times 48$& 0.1245& 0.12820 &  300& 29\\
$\ED$&$48 \times 24^2 \times 48$& 0.1245& 0.12820 &  270& 37\\
\cmidrule[0pt]{1-6}
$\EEA$&$16\times16^2 \times 32$& 0.1215& 0.12838  & 230 & 16\\
$\EEB$&$24\times 24^2 \times 64$& 0.1215& 0.12838 & 450& 23 \\
$\EEC$&$28\times 24^2 \times 64$& 0.1215& 0.12838 & 670& 33\\
$\EED$&$32\times 24^2 \times 64$& 0.1215& 0.12838 & 500& 37\\
\bottomrule
\end{tabular*}
\caption{Details of the lattice ensembles used in this work. $N_c$ and $N_s$ label the number of configurations and number of sources on each configuration, respectively. The top four ensembles correspond to $m_\pi=306(1)\MeV$ and the bottom four $m_\pi=227(2)\MeV$.}
\label{tab:ensembles}
\end{table}

To reduce the statistical uncertainties we computed quark propagators at multiple point sources for 
each configuration. Since the presence of the Dirichlet walls breaks translational symmetry in the
$x$ and $t$ directions, the point sources have to be picked carefully; they were displaced with 
respect to each other using translations in the $y$ and $z$ directions, which have periodic boundary conditions.  

\begin{table}[b]
\begin{tabular*}{0.8\columnwidth}{@{\extracolsep{\stretch{1}}}cccc@{}}
\toprule
Ensemble & Pion&Kaon&Neutron\\
\midrule                                                                     
$\EA$ & [10, 19]& [10, 19] & [8, 21]\\
$\EB$ &  [14, 30]& [14, 30] & [8, 21]\\
$\EC$ & [13, 30]& [13, 30] & [9, 21]\\
$\ED$ & [14, 30]& [14, 30] & [8, 21]\\
\cmidrule[0pt]{2-4}
$\EEA$ &  [10, 19]& [10, 19] & [9, 21]\\   
$\EEB$ & [15, 36]& [15, 37] & [9, 21]\\   
$\EEC$ &  [15, 37]& [15, 37] & [10, 21]\\   
$\EED$ &  [15, 37]& [15, 36] & [9, 21]\\   
\bottomrule
\end{tabular*}
\caption{Fit ranges used in extracting the energy shifts for the pion, kaon, and neutron.}
\label{tab:fitwindow}
\end{table}

\begin{figure*}[t]
\centering
\includegraphics[width=0.48\textwidth]{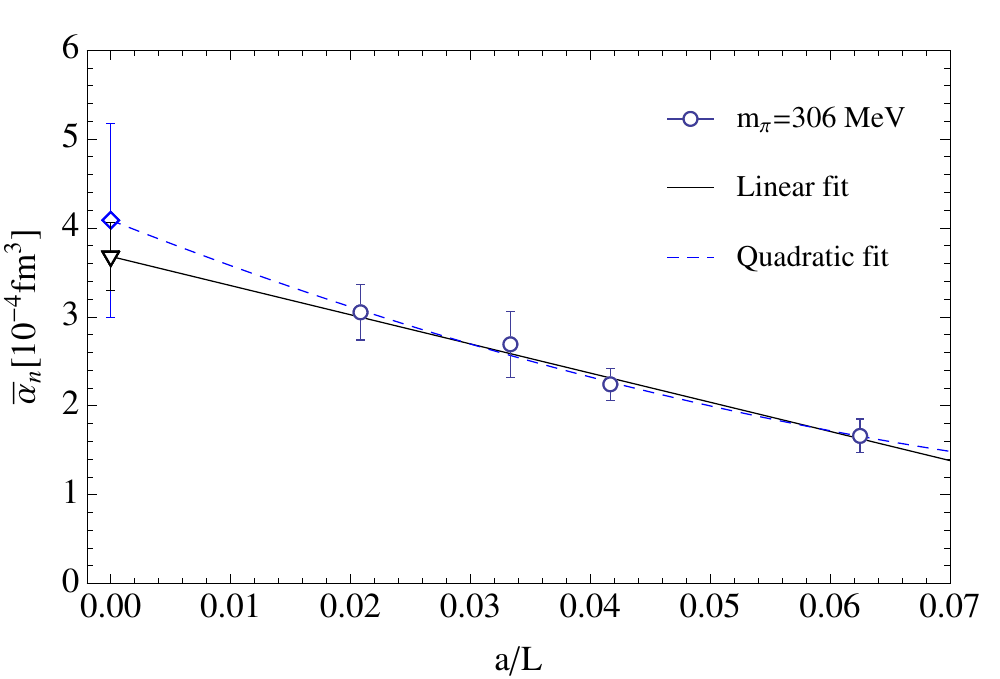}
\includegraphics[width=0.48\textwidth]{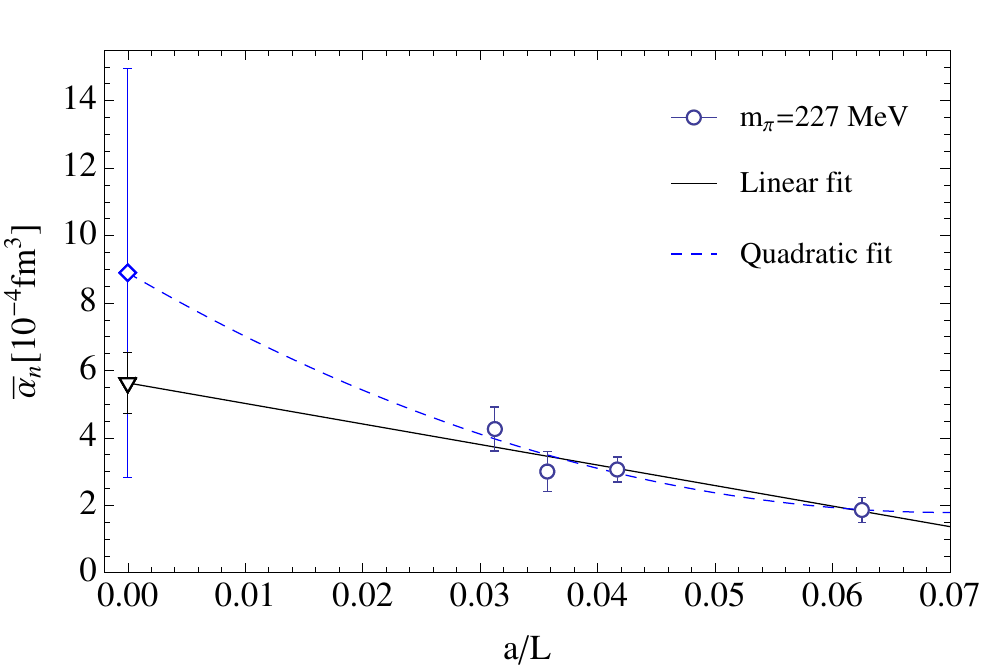}
\caption{Infinite volume extrapolation for neutron polarizability. The left panel shows our results for the 
$m_{\pi}=306$ MeV ensembles and the right panel for the $m_{\pi}=227$ MeV ensembles. On each plot we overlay 
the infinite volume extrapolations using a linear (solid line) or quadratic (dashed line) fit.}
\label{plot:nvolplot1}
\end{figure*}

To determine the appropriate time window to fit the correlation functions, we varied the start 
time, $t_\text{min}$ and kept the maximum fit time fixed. For each case we performed a fit and 
extracted the hadron's energy shift, $\delta E$, and the associated $\chi^2$/dof. 
Following the procedure discussed in~\cite{Lujan:2014kia}, we choose the largest 
fit window that produces a good quality fit. The fit windows used for each of the hadrons
studied in this paper are listed in Table~\ref{tab:fitwindow}. 
The computed values for the polarizability of the three hadrons are presented in Tables~\ref{tab:fitdata}. 
In the same table we include the energy shifts due to the field, the energies measured in the absence of the
field with Dirichlet boundary conditions, and the masses as extracted using periodic boundary conditions.  

Since we use dozens of point sources for each ensemble, and for each point source we 
need to compute the quark propagator for five different couplings to the background electric 
field, we have to compute hundreds of inversions for each configuration. To compute these 
efficiently, we use our implementation of a multi-GPU Dslash operator~\cite{Alexandru:2011sc} 
and an efficient multi-mass inverter~\cite{Alexandru:2011ee}.


\section{Volume Dependence Analysis} \label{sec:voldependence}

Finite volume corrections have been estimated using $\chi$PT.  For periodic boundary conditions these effects were calculated for electric polarizabilities~\cite{Detmold:2006vu} and magnetic polarizabilities~\cite{Hall:2013dva}. At $m_{\pi}$ around 250 MeV and $L=3\,\fm$ it was estimated that the correction to the neutron polarizability is about 7\%~\cite{Detmold:2006vu}.  For Dirichlet boundary conditions used in this work, no direct $\chi$PT predictions are available.  The only estimate comes from sigma model studies of the chiral condensate in the presence of hard walls~\cite{Tiburzi:2013vza}. This choice of boundary conditions is expected to introduce larger finite volume effects that are expected to vanish algebraically with $1/L_x$ in the infinite volume limit.  The expectation is based on the idea that the corrections are mainly driven by the hadron momentum $\pi/L_x$.  To thoroughly analyze the volume dependence we performed our calculations on four different lattice sizes for both pion masses.

\begin{table}[b]
\begin{tabular*}{\columnwidth}{@{\extracolsep{\stretch{1}}}l*{7}c@{}}
\toprule
\multicolumn{1}{c}{}& \multicolumn{3}{c}{ $306\MeV$ }&\phantom{aa} &\multicolumn{3}{c}{ $227\MeV$ }\\
 &$\bar\alpha_n$ & $\chi^2$& AIC & &$\bar\alpha_n$&$\chi^2$ & AIC\\
 \cmidrule{2-4}\cmidrule{6-8}
    Constant & 2.18(11)&17.4& 19.40&&  2.77(22)&11.76 &13.76 \\
    Linear &3.67(38) &0.298 & 4.30 &&  5.62(91) &1.28 &5.28 \\
    Quadratic &4.1(1.1) &0.141&6.14&&  8.9(6.1)  &0.99 &6.99\\
\bottomrule
\end{tabular*}
\caption{Infinite volume extrapolation results for the neutron with three different fit models.
The polarizabilities $\bar\alpha_n$ are reported in units of $10^{-4}\fm^3$.}
\label{tab:ninfres}
\end{table}

Since we do not know the analytical form for the finite volume effects, we fit the polarizability as a function of $1/L$ to three different models: constant, linear, and quadratic. We cannot go beyond the quadratic since we only have four different lattice sizes. To determine which model fits the data best we compute the $\chi^2$ to gauge the overall goodness of the fit.  In conjunction with the goodness of fit criteria we  use Akaike Information Criterion (AIC)~\cite{Akaike:1974}, which measures the relative quality between different statistical models and helps in determining whether or not a model is overfitting the data. The AIC value is given by
\begin{equation}
\mbox{AIC} = 2k + \chi^2,
\end{equation}
where $k$ is the number of parameters in the model.
For a given fit model we will sum the values of the AIC for both pion masses.  The model with the smallest AIC value will be our fit model used subsequently.

%
%

\subsection{The Neutron}\label{sec:neutrondisc}

The extrapolation results for the neutron polariability are tabulated in Table~\ref{tab:ninfres}.  Figure~\ref{plot:nvolplot1} plots our polarizability results along with the linear and quadratic fits which had the smallest values for the AIC and good $\chi^2$ values. Both the linear and quadratic models produced consistent results. However, the linear model produces a smaller value for the AIC which indicates that the quadratic model may be overfitting the data.  We will use the linear infinite volume results when discussing the chiral behavior of the neutron.

\begin{figure}
\centering
\includegraphics[width=0.95\columnwidth]{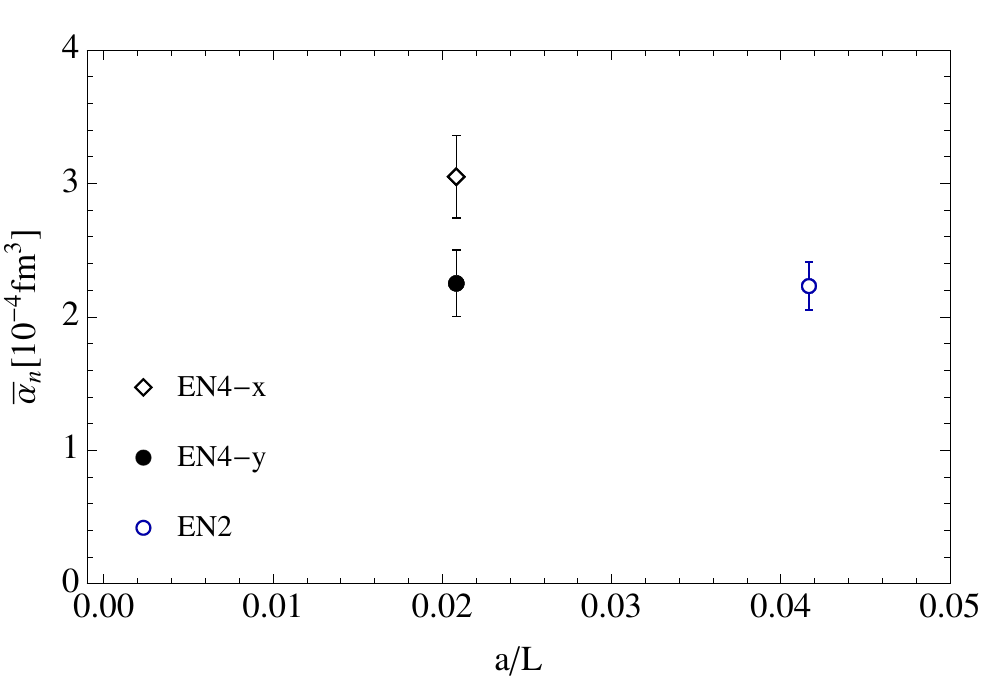}
\caption{Comparison of the $\EB$ spatial lattice ($24\times 24^2$) with the electric field in the 
$x$-direction to the results of the $\ED$ spatial lattice ($48\times24^2$) with the electric field 
in both $x$- and $y$-directions. The results confirm that the finite size effects associated with the
directions transverse to the electric field are negligible.}
\label{plot:nvolplot2}
\end{figure}

The volume dependence analysis assumes that the finite size effects, due to the electric field, 
are determined by the size of the lattice parallel to the applied field~(which is in $x$-direction for this work). 
To verify this, we take our $\ED$ lattice which has the spatial dimension $48\times24^2$ and place the 
electric field along the $y$-direction which has only 24 lattice units. We choose this ensemble because the 
difference in the $x$ and $y$ directions are the largest which gives us the best comparison. We expect our 
results to be comparable to the results of the $\EB$ ensemble which has the  spatial dimension $24\times 24^2$.  
We find $\bar{\alpha}_n = 2.25(25) \times 10^{-4}\, \fm^3$, which is statistically equivalent to the polarizability 
for the $\EB$ and significantly different from the case where we place the field along the $N_x=48$ direction. 
Fig.~\ref{plot:nvolplot2} displays the comparison. We conclude that the finite volume effects associated with
the directions perpendicular to the field are negligible.

\subsection{Pion and Kaon}

\begin{figure*}
\centering
\includegraphics[width=0.48\textwidth]{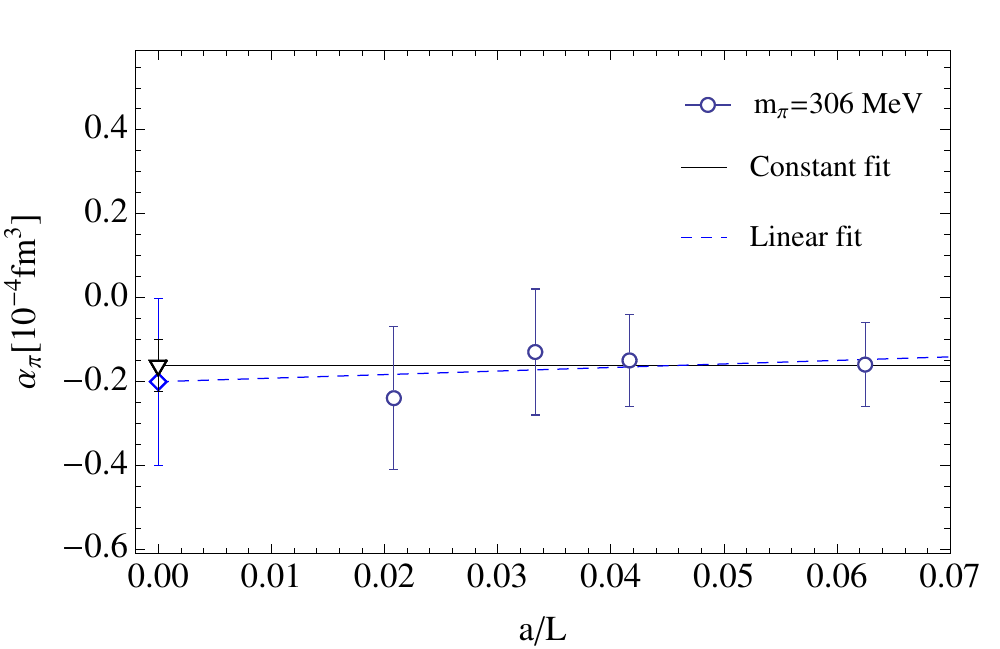}
\includegraphics[width=0.48\textwidth]{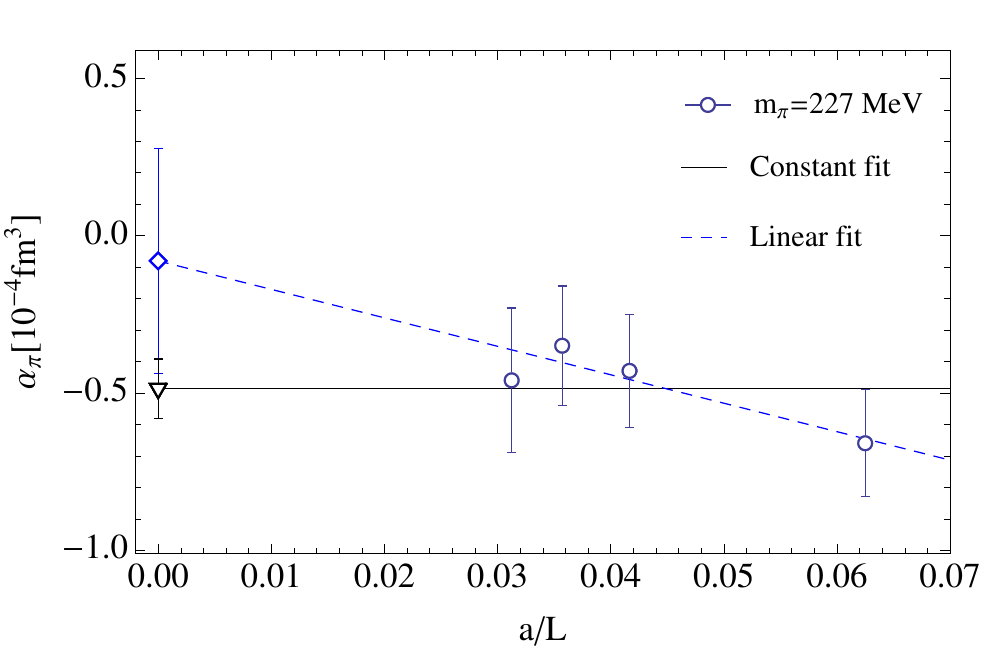}
\includegraphics[width=0.48\textwidth]{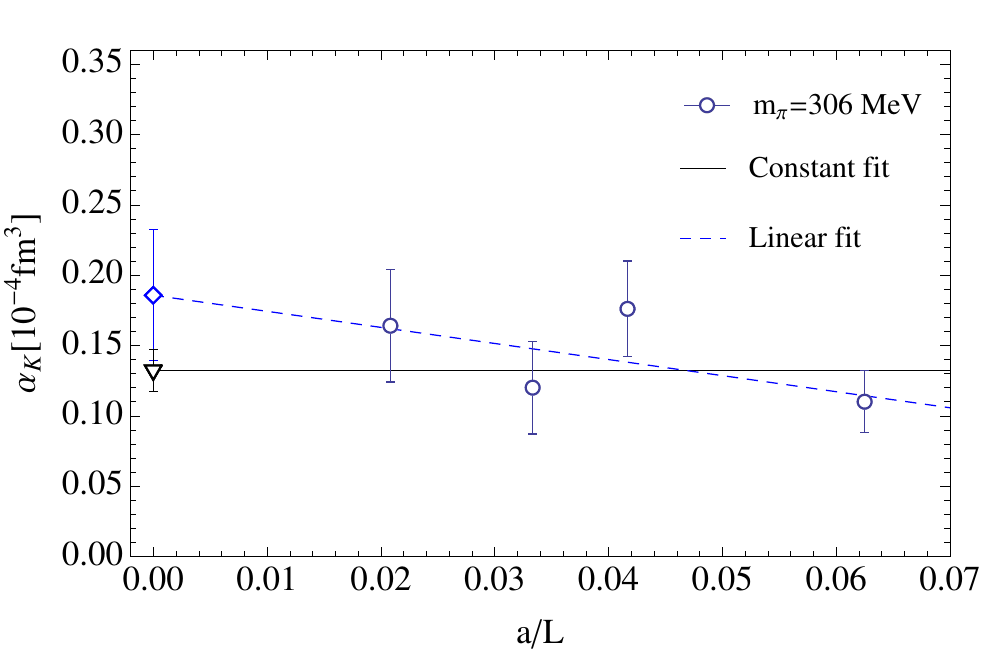}
\includegraphics[width=0.48\textwidth]{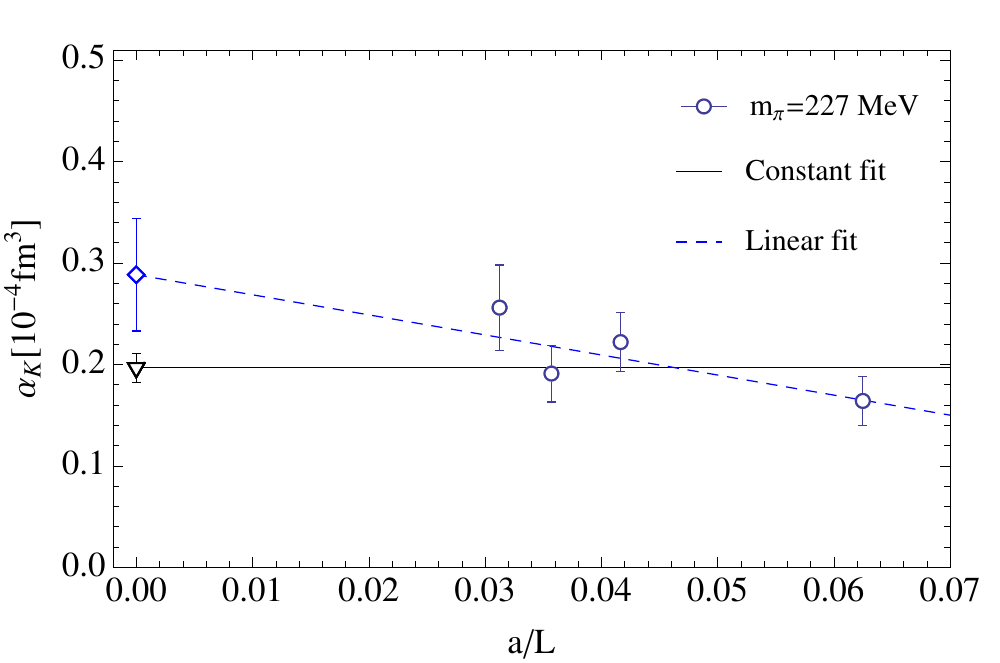}
\caption{Infinite volume extrapolation for pion~(top) and kaon~(bottom) polarizability
for $m_\pi=306\MeV$~(left) and $m_\pi=227\MeV$~(right).
The two lines are infinite volume extrapolations using a constant (solid line) or a linear (dashed line) fit.}
\label{plot:pvolplot1}
\end{figure*}
The volume dependence analysis for the pion and kaon proceeds in the same way as the neutron. 
Fig.~\ref{plot:pvolplot1} shows our extracted polarizabilities as a function of $1/L$ for the pion~(top plots) 
and kaon~(bottom plots). We also plot the results of the constant and linear extrapolations which were the two 
models with the smallest values for the AIC. 
The results of the extrapolation are tabulated in Tables~\ref{tab:pinfres}~and~\ref{tab:kinfres}. For the pion 
we find that the constant fit model gives the smallest combined AIC values. For the kaon at $m_\pi=306 \MeV$ the 
constant model gives a smaller value of the AIC than the linear model. However, the combined result for both pion 
masses---the AIC coefficient for the combined fit is the sum of coefficients for the individual fits---are smaller 
for the linear model. We therefore use the linear model for the kaon.

\begin{table}[b]
\begin{tabular*}{\columnwidth}{@{\extracolsep{\stretch{1}}}l*{7}c@{}}
\toprule
\multicolumn{1}{c}{}& \multicolumn{3}{c}{ $306\MeV$ }&\phantom{aa} &\multicolumn{3}{c}{ $227\MeV$ }\\
 &$\alpha_\pi$ & $\chi^2$& AIC & &$\alpha_\pi$&$\chi^2$ & AIC\\
 \cmidrule{2-4}\cmidrule{6-8}
    Constant & -0.16(6)&0.27& 2.27  &&  -0.486(94)&  1.67& 3.67 \\
    Linear &-0.20(20)&0.23 & 4.23   &&  -0.08(36)& 0.29& 4.29 \\
    Quadratic &-0.44(61)&0.06 &6.06 &&  -1.1(2.5)&0.12 &6.12\\
\bottomrule
\end{tabular*}
\caption{Infinite volume extrapolation results for the pion with three different fit models.
The polarizabilities $\alpha_\pi$ are reported in units of $10^{-4}\fm^3$.}
\label{tab:pinfres}
\end{table} 
  
%
%
%
%
\begin{table}[b]
\begin{tabular*}{\columnwidth}{@{\extracolsep{\stretch{1}}}l*{7}c@{}}
\toprule
\multicolumn{1}{c}{}& \multicolumn{3}{c}{ $306\MeV$ }&\phantom{aa} &\multicolumn{3}{c}{ $227\MeV$ }\\
 &$\alpha_{K^0}$ & $\chi^2$& AIC & &$\alpha_{K^0}$&$\chi^2$ & AIC\\
 \cmidrule{2-4}\cmidrule{6-8}
    Constant & 0.132(15)& 3.45& 5.45&&  0.197(14)  & 4.65& 6.65 \\
    Linear &0.186(47) &1.98 &5.98 &&  0.289(55) & 1.71& 5.71 \\
    Quadratic &0.12(15)&1.80 & 7.8&&  0.29(42) &1.71 & 7.71\\
\bottomrule
\end{tabular*}
\caption{Infinite volume extrapolation results for the kaon with three different fit models.
The polarizabilities $\alpha_{K^0}$ are reported in units of $10^{-4}\fm^3$.}
\label{tab:kinfres}
\end{table}

 



\section{Discussion} \label{sec:discussion}

In this section we discuss our infinite volume results for the polarizability of the neutral pion and kaon and neutron 
in the context of other calculations on the lattice, chiral perturbation theory ($\chi$PT), and experiment.

\begin{figure}
\includegraphics[width=0.97\columnwidth]{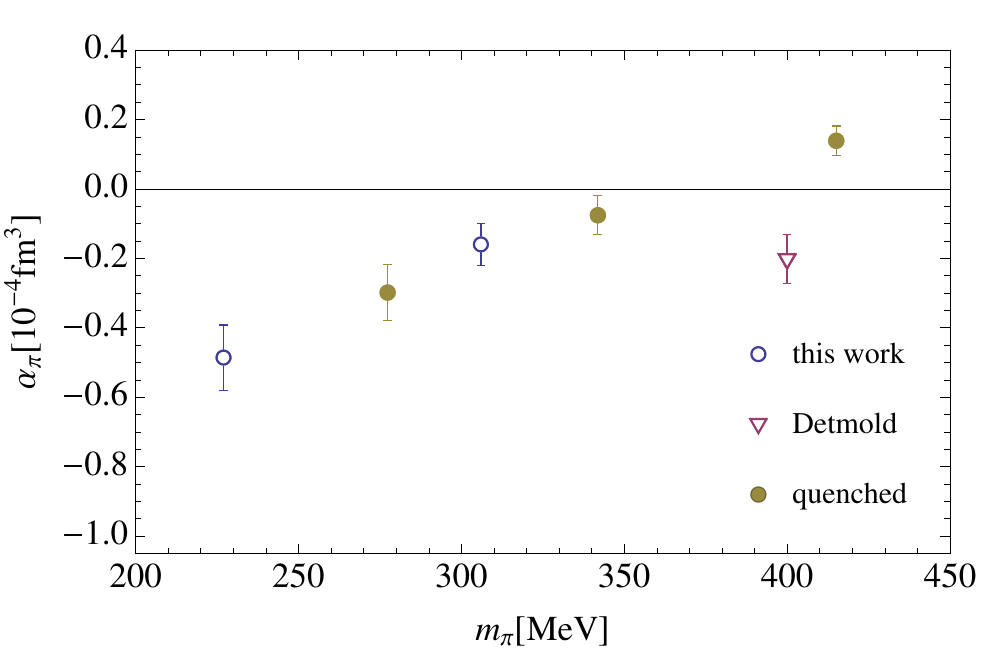}\\
\hskip3mm\includegraphics[width=0.945\columnwidth]{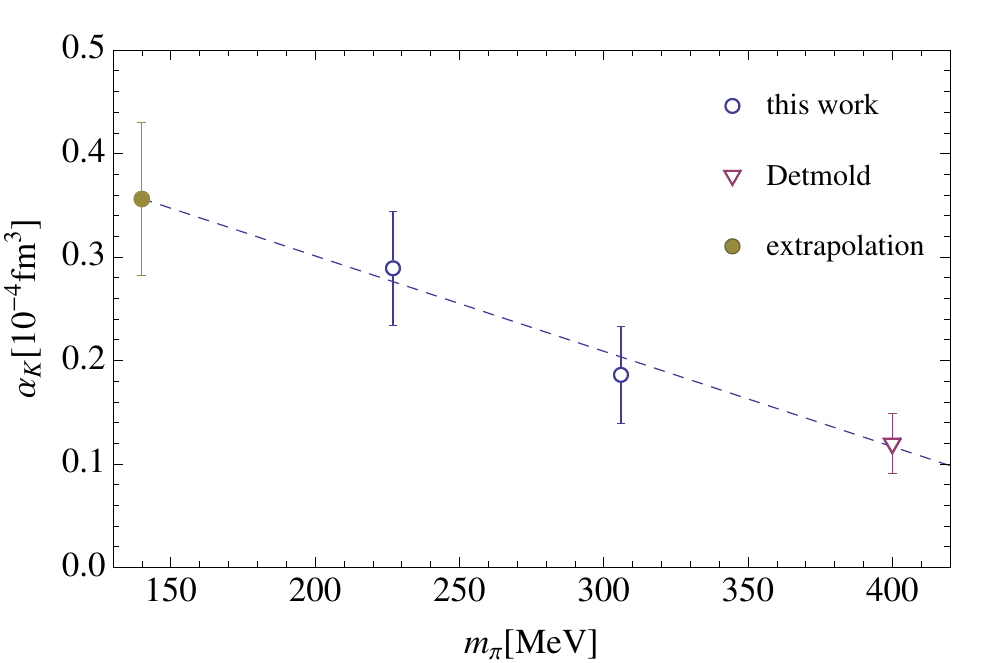}
\caption{Top: Neutral pion polarizability as a function of the quark mass. The circles are 
quenched results found in \cite{Alexandru:2010dx} and the triangle is the value determined 
in~\cite{Detmold:2009dx}. Bottom: Neutral kaon polarizability
along with a chiral extrapolation which includes the value determined in~\cite{Detmold:2009dx}. }
\label{plot:chiralpionkaon}
\end{figure}

For the neutral pion, the results are summarized in the top panel of Fig.~\ref{plot:chiralpionkaon}. 
In addition to our dynamical results, 
we also show the infinite volume results from our quenched study~\cite{Alexandru:2010dx}. Since the
finite volume corrections are insignificant, the conclusions from our recent study~\cite{Lujan:2014kia} are
unchanged: the polarizability depends very little on the mass of the sea quarks, but it changes as we vary
the mass of the valence quarks. The puzzling feature persists: the neutral pion polarizability becomes 
negative for $m_\pi \approx 350\MeV$,  and its magnitude increases as we approach the physical point. 
The negative trend was also observed by Detmold {\it et al.}~\cite{Detmold:2009dx} 
as indicated by their result (the blue triangle) at $m_{\pi} = 400\MeV$ on the same plot. 
It was pointed out in Ref.~\cite{Detmold:2009dx} that the negative value is inconsistent with expectations from 
$\chi$PT when only the connected part of the correlator is included, as is the case in both lattice calculations.
It was speculated in Ref.~\cite{Detmold:2009dx} that the negative value could arise due to finite volume effects. 
Our infinite volume results demonstrate that it does not seem to be the case. Other effects might be at play.
One possible contribution to this discrepancy is the fact that the sea quarks are electrically neutral in these studies.
We have investigated the effects of charging the sea quarks~\cite{Freeman:2014kka}, and our initial results hint 
at this scenario: we found that the neutral pion polarizability changes sign as we charge the sea quarks, 
albeit still with large statistical errors. Efforts are under way to reduce the errors. 
Note that there remains a bit of disagreement between the trend suggested by our results
and the data from Ref.~\cite{Detmold:2009dx}.  
It is not clear whether the disagreement is due to finite volume effects, 
discretization errors (we use different actions), or statistical fluctuations.
Recently it was suggested that discretization errors present for Wilson-type fermions used in this study and the
other mentioned above might be responsible for these puzzles~\cite{Bali:2015vua}. The background field changes the value of the
additive mass renormalization and this might lead to energy shifts in hadron mass unrelated to polarizabilities.
A continuum limit study is required to determine whether this effect is large enough to explain these puzzles.
Ultimately, the disconnected contribution must also be included to complete the picture for the neutral pion polarizability.

For neutral kaon our results are presented in the bottom panel of Fig.~\ref{plot:chiralpionkaon}. 
In contrast to the pion case, neutral kaon has a stronger dependence on the sea quarks.
In our previous study we performed a chiral extrapolation and we found  $\alpha_{K^0} = 0.269(43)$ in units 
of $10^{-4} \fm^3$~\cite{Lujan:2014kia}. We perform the same chiral extrapolation using a linear ansatz in 
$m_{\pi}$ but now using our infinite volume values. We include the value determined by 
Detmold et al.\ \cite{Detmold:2009dx} since the finite volume corrections decrease with increasing 
$m_\pi$ and we expect it to be negligible at $400\MeV$. 
We find $\alpha_{K^0}= 0.356(74)\times10^{-4} \fm^3$, only slightly higher than 
the finite volume value,  suggesting that the finite volume corrections are small for the kaon. 
The relative smallness of the neutral kaon polarizability is consistent with $\chi$PT 
which predicts a vanishing value at the one-loop level, even with electrically neutral sea quarks~\cite{Guerrero:1997rd}.

\def\chiptg{HB$\chi$PT-NNLO}
\def\chiptp{B$\chi$PT-NLO}

\begin{figure}
\centering
\includegraphics[width=1.03\columnwidth]{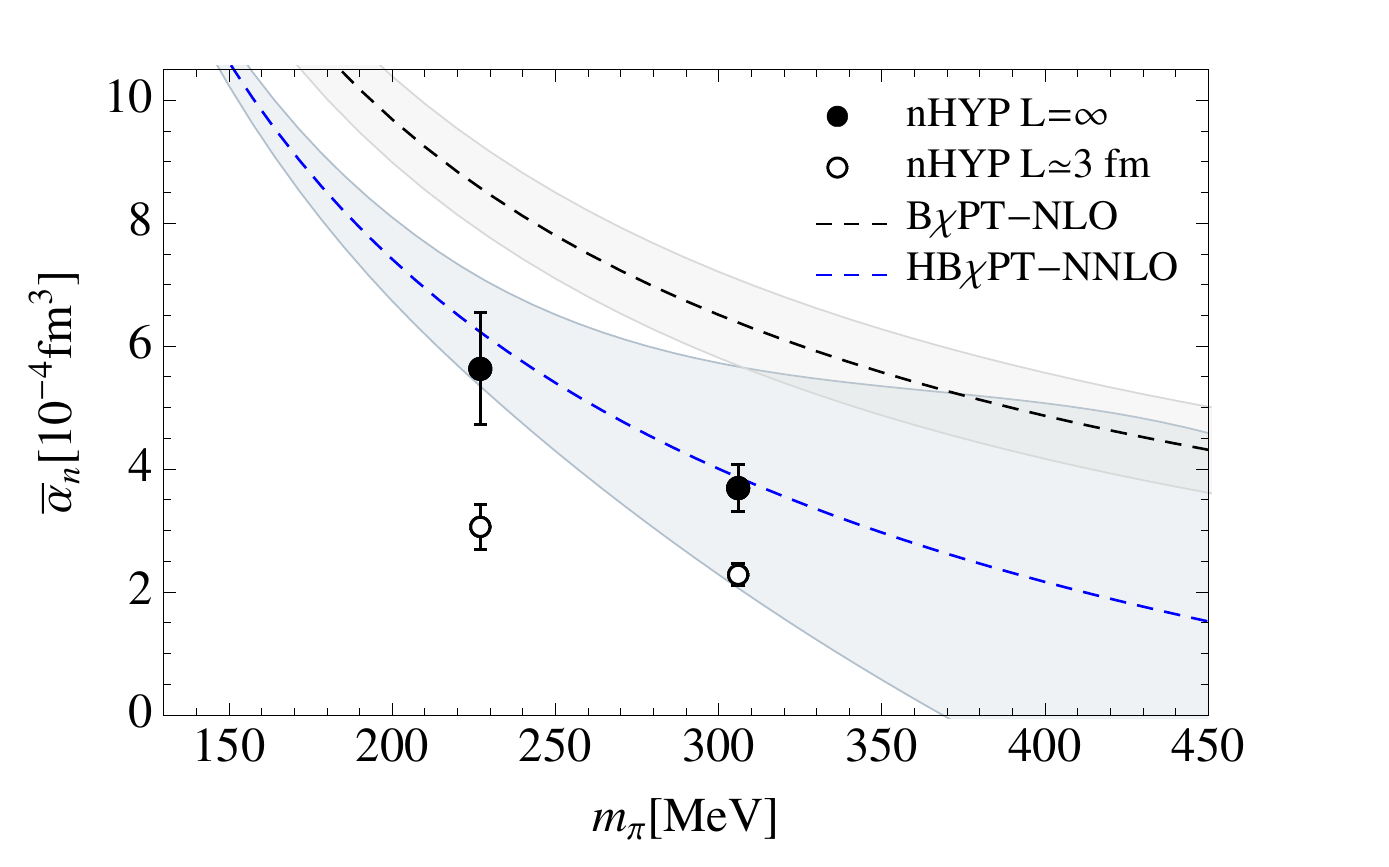}
\includegraphics[width=1.03\columnwidth]{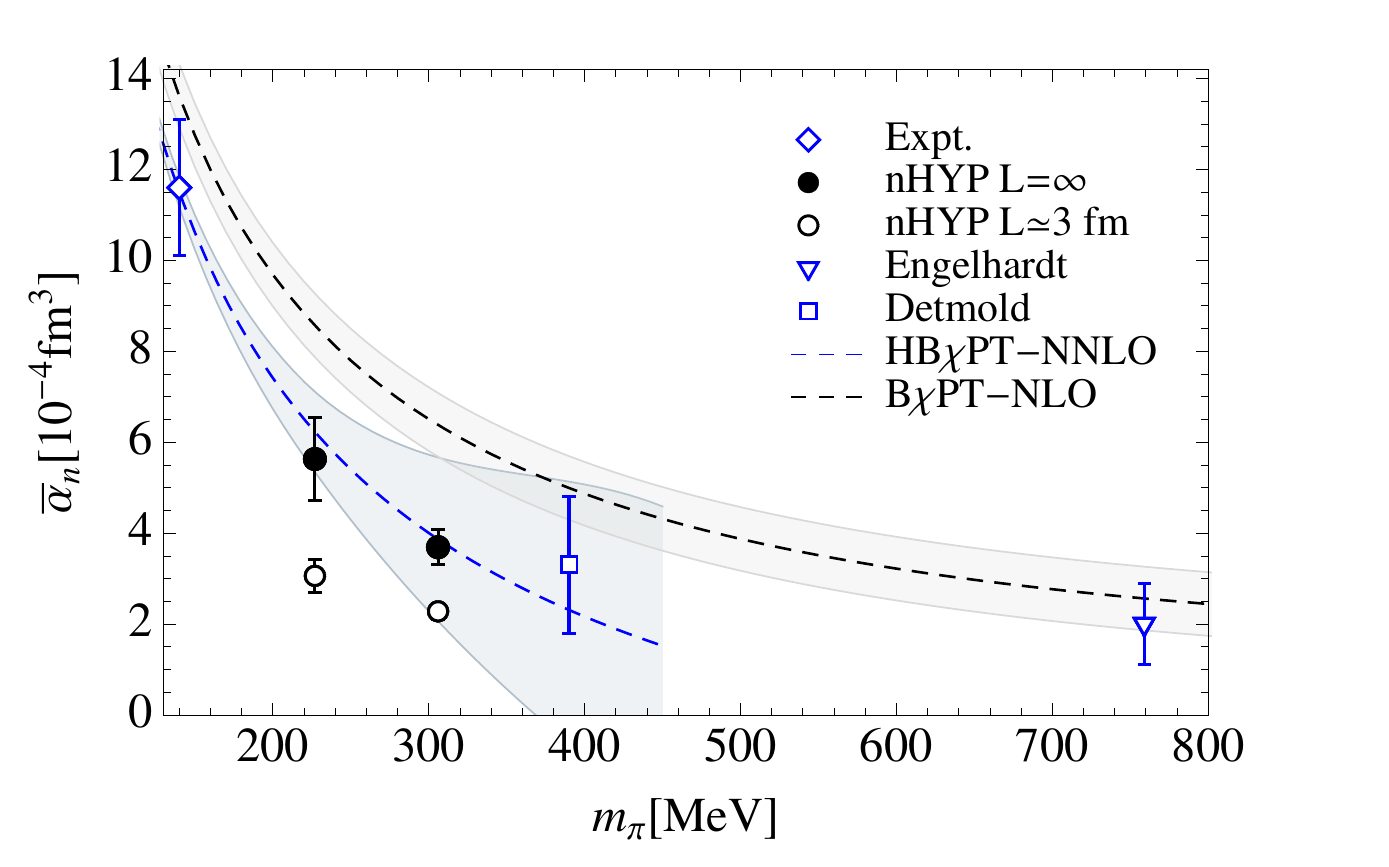}
\caption{Top: Neutron polarizability as a function of quark mass.
The black empty circles are our finite volume results presented in~\cite{Lujan:2014kia} 
and the full circles are our infinite volume results. The dashed lines are two different 
$\chi$PT calculations: \chiptg~\cite{Griesshammer:2015ahu} and \chiptp~\cite{Lensky:2009uv}. 
The uncertainties in the curves are indicated by the shaded regions. 
Bottom: Comparison with the experimental value and two other lattice calculations~\cite{Engelhardt:2007ub} 
and~\cite{Detmold:2010ts}. }
\label{plot:nvolplotchiral1}
\end{figure}

We turn the discussion now to the neutron. In the top panel of Fig.~\ref{plot:nvolplotchiral1} we display
the neutron electric polarizability as a function of $m_{\pi}$.  
We compare our results to two different $\chi$PT curves: a N${}^2$LO calculation using a nonrelativistic 
form for some of the propagators (\chiptg)~\cite{Griesshammer:2015ahu}, 
and a NLO result that uses relativistic propagators (\chiptp)~\cite{Lensky:2009uv}. 
We see that the value for $m_\pi=227\MeV$ computed on a box with $L\approx 3\fm$ disagrees with both curves. 
After correcting for the finite volume effects, our results agree very well with the \chiptg\ curve.
In the right panel of Fig.~\ref{plot:nvolplotchiral1} we show our results together with the experimental 
value and compare them with two other lattice results~\cite{Detmold:2010ts,Engelhardt:2007ub} obtained on finite lattices. 
We see that our results have significantly smaller statistical errors
even though they are computed using smaller pion masses and they are extrapolated to infinite volume.

\begin{figure}[h]
\centering
\includegraphics[width=\columnwidth]{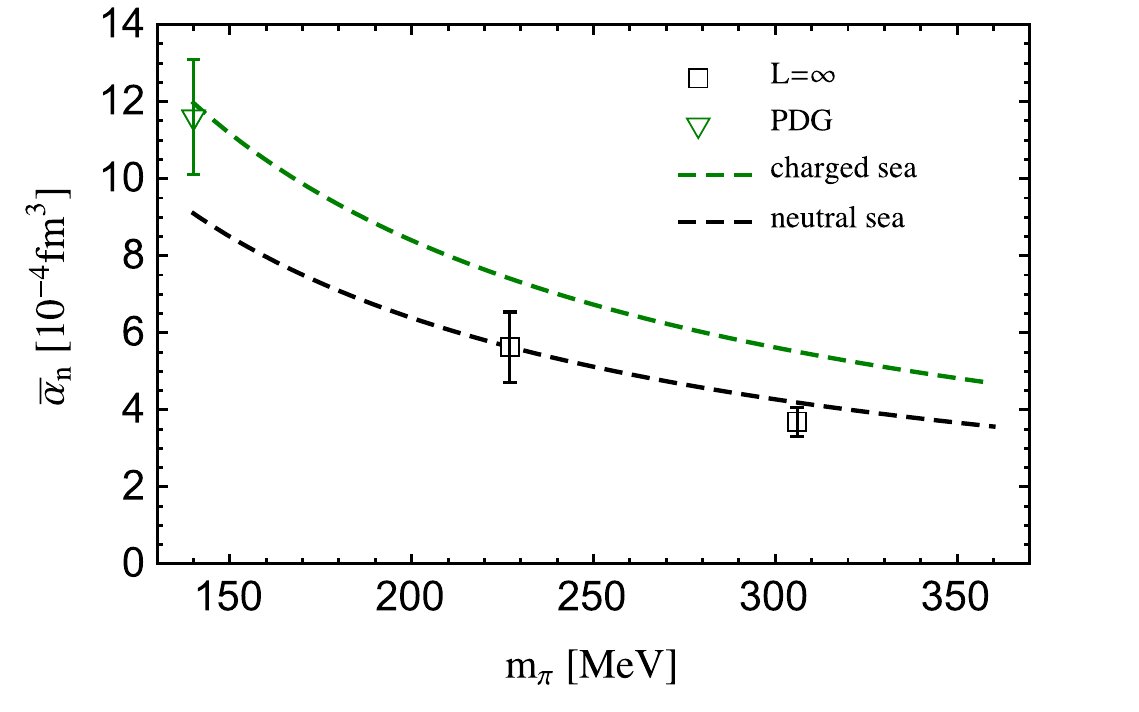}
\caption{Expected sea quark charging effects in the neutron polarizability. 
Our infinite volume results are plotted along with the $\chi$PT predictions from~\cite{Detmold:2006vu} 
with neutral and charged sea quarks.}
\label{plot:nvplotcharge2}
\end{figure}

This analysis demonstrates that finite volume effects are very important for neutron polarizability.
We expect that any other systematic effects are small and that the calculation, 
for the pion masses used in this study, is nearly complete. 
The discretization effects are expected to be of the order of one percent as experience with similar
actions indicates~\cite{Durr:2010aw}. The only remaining significant systematic error comes 
from neglecting the charge of the sea quarks. For the $\EB$ ensemble, the correction was already 
computed~\cite{Freeman:2012cy}. The effect was found to be small, similar to the size of statistical errors.
This is also supported by a partially quenched $\chi$PT calculation~\cite{Detmold:2006vu}: using the
formulas derived in that paper, we find that for $140\MeV \leq m_\pi \leq 300\MeV$, neutron electric
polarizability increases by a value of 1.5 to 2 in units of $10^{-4}\fm^3$, when the sea quark charges are turned on.
This prediction is shown in Fig.~\ref{plot:nvplotcharge2}. To produce these curves we used the parameters 
suggested in the paper, but we had to set $|g_{N\Delta}| = 0.25$ (a value outside the expected range) to 
make the ``charged" curve go through the experimental point. Our results, which were derived using neutral 
sea quarks, agree very well with the ``neutral" curve.

Before we conclude, we would like to discuss the systematic error
associated with the choice for the fitting window. To gauge this 
error, we extract the energy shift using two other fit windows---one shifted by one unit 
in positive time direction and one shifted in the negative 
direction---and repeated the analysis. For the infinite volume 
extrapolations we use a linear fit for neutron and kaon, and a constant fit for 
the pion. The systematic error quoted here is the standard deviation of the final
results extracted using our three fit windows. For the neutron we have 
$\bar\alpha_n=3.67(38)(27)$ and $\bar\alpha_n=5.62(91)(89)$ for $m_\pi=306\MeV$
and $227\MeV$ respectively. Similarly, for neutral pion we have 
$\alpha_{\pi}=-0.16(6)(6)$ and $\alpha_{\pi}=-0.486(94)(46)$ and
for neutral kaon $\alpha_{K^0}=0.186(47)(29)$ and $\alpha_{K^0}=0.289(55)(52)$.
The polarizability for neutral kaon at the physical point is $\alpha_{K^0}=0.356(74)(46)$.
All the results here are presented in natural units for hadron polarizabilities of
$10^{-4}\fm^3$, with the first error being stochastic and the second the systematic
due to fit window. Note that this systematic is smaller or comparable with the 
stochastic error.

\section{Conclusion} 
\label{sec:conclusion}

We have analyzed the volume dependence of the electric polarizability $\alpha$ 
for the neutral pion, neutral kaon, and neutron on four different lattice volumes 
at two light quark masses corresponding to pion masses of 306 and 227~MeV, in the
mass region where chiral perturbation theory predictions are most likely reliable.
The novel aspect of this calculation is that it is the first systematic study 
of finite volume effects on polarizability in the presence of
Dirichlet boundaries. These boundary conditions allow for very weak electric fields 
in order to avoid a possible vacuum instability. 
We also estimate the effects of charging the sea quarks. 

For the neutral pion, our results confirm  that the negative trend in the polarizability is not due to finite volume effects. 
Rather, preliminary results indicate that the behavior is most likely due to the neglecting of the charge in the sea quarks. 
To compare with experiment, the disconnected contribution to the neutral pion polarizability will have to be included.

For the neutral kaon, we performed a similar chiral extrapolation to the physical point as was done 
in~\cite{Lujan:2014kia} but now using the infinite volume extrapolations for $\alpha_{K^0}$. We find 
$\alpha_{K^0}(m_\pi^\text{phys})= 0.356(74)\times10^{-4} \fm^3$ which is only slightly higher than the value determined on box 
sizes $L\simeq 3 \fm$. This indicates that the volume effects for the kaon polarizability are relatively mild. 

For the neutron we find that the finite volume corrections are important. After removing them, our results are now in 
excellent agreement with predictions from chiral perturbation theory. 
We have not yet performed a chiral extrapolation for the neutron
since we still need to include the corrections due to the interactions
between the sea-quarks and the background field.
We are currently investigating the best method to do the extrapolation using input from $\chi$PT. 

We are in the process of including the effect of charged sea quarks in the analysis for all our ensembles. Along with the infinite-volume extrapolation done here, this is part of our program geared toward determining the polarizabilities at the physical point.

\begin{acknowledgements}
We thank Andr\'{e} Walker-Loud, Vladimir Pascalutsa, and Harald Grie$\ss$hammer for discussions and 
correspondence related to this project.
The computations were carried out on a variety of GPU-based supercomputers, including the GWU IMPACT collaboration machines 
and Colonial One cluster, and USQCD resources at Jefferson Lab and Fermilab. This work is supported in part by the 
NSF CAREER grant PHY-1151648, the U.S. Department of Energy grant DE-FG02-95ER40907, and the ARCS foundation.
\end{acknowledgements}

\begin{table*}
\begin{tabular*}{0.95\textwidth}{@{\extracolsep{\stretch{1}}}*{12}c@{}}
\toprule
 &Hadron& \phantom{a}& $\EA$& $\EB$& $\EC$& $\ED$&\phantom{a}& $\EEA$&$\EEB$ &$\EEC$&$\EED$\\

 \cmidrule{4-7}\cmidrule{9-12}
\multirow{3}*{\parbox[t]{1.25cm}{$\alpha$\\$[10^{-4}\fm^3]$}}
    &\multirow{1}*{$\pi$}&& -0.160(10) & -0.15(11)& -0.13(15)&-0.24(0.17) && -0.66(17) &-0.43(18)&  -0.35(19)& -0.46(23)\\
    &\multirow{1}*{$K$} &&0.110(22) & 0.176(34)& 0.120(33) & 0.164(40) && 0.164(24) &0.222(29) &0.191(28) &0.256(42) \\                                                               
    &\multirow{1}*{$n$} && 1.66(19)&2.23(18) & 2.69(37) &3.05(31) && 1.86(38) &3.06(37) & 3.00(59)& 4.26(65)\\                                                                                               
\midrule[0pt]
\multirow{3}*{\parbox[t]{1.25cm}{$a\delta E$\\$[\times10^{-8}]$}}
&    \multirow{1}*{$\pi$}&& -3.15(2.00) & -3.69(2.77)& -3.40(4.13)& -7.13(5.09) && -11.92(3.07) & -10.01(4.24) & -9.0(4.8) &-12.50(6.27)\\
&    \multirow{1}*{$K$} &&2.83(57)& 5.14(99) & 3.47(1.00)&5.22(1.28) && 4.84(72) & 7.14(94) & 6.22(91)&8.82(1.47)\\                                                            
&    \multirow{1}*{$n$} &&33.1(5.0) &53.4(5.4)& 72.6(12) &78.9(9.1) && 41.65(11.2) & 86.2(12.5) & 87.8(20.0)&125.3(20.7)\\   
\midrule[0pt]
\multirow{3}*{\parbox[t]{1.25cm}{$a E$\\}}
&    \multirow{1}*{$\pi$}&&0.322(35) & 0.251(9)& 0.2362(9)& 0.2084(9) && 0.276(6)& 0.207(1)& 0.184(1)&0.176(1) \\
&    \multirow{1}*{$K$}&&0.401(2)& 0.3515(8) & 0.3566(8) & 0.3241(7) && 0.433(1) & 0.3952(6) & 0.392(2) & 0.3711(10) \\                                                            
&    \multirow{1}*{$n$} && 0.768(16) & 0.696(9) & 0.658(10)  & 0.689(2) && 0.710(6) & 0.634(4) &0.610(4) &0.619(7) \\
\midrule[0pt]
\multirow{3}*{\parbox[t]{1.25cm}{$a m$\\}}
&    \multirow{1}*{$\pi$}&&0.1986(22) &0.1932(7)&0.1934(8) & 0.1938(8) && 0.145(3) & 0.140(1)& 0.138(1)&0.1391(8)\\
&    \multirow{1}*{$K$} &&0.3235(15) & 0.3220(7)&  0.3228(8)& 0.3229(7) && 0.372(1) & 0.3698(6) &0.371(2) &0.372(1)\\                                                            
&    \multirow{1}*{$n$} &&0.642(11) &0.644(6)& 0.657(8) & 0.647(4) && 0.622(20)& 0.618(13)& 0.620(23)&0.60(3)\\
\bottomrule
\end{tabular*}
\caption{Electric polarizabilities, energy shifts due to the field, energies computed with no external field, 
and masses extracted from boxes with periodic boundary conditions for the pion, kaon, and neutron for 
the 8 ensembles used in this study.}
\label{tab:fitdata}
\end{table*}

\bibliographystyle{jhep}
\bibliography{my-inf-vol-ref.bib}
\end{document}